\documentclass[aps,prl,twocolumn,showpacs,superscriptaddress,unsortedaddress,groupedaddress,altaffillsymbol]{revtex4}
\usepackage{braket}
\usepackage{upgreek}
\usepackage{graphicx}
\usepackage{amsmath}
\usepackage{bm}        
\usepackage{amssymb}   

\begin{document}

\title{High Speed, High Fidelity Detection of an Atomic Hyperfine Qubit}
\date{\today}

\author{Rachel Noek}
\author{Geert Vrijsen}
\author{Daniel Gaultney}
\author{Emily Mount}
\affiliation{Electrical and Computer Engineering Department, Duke University, Durham, N.C. 27708, USA}
\author{Taehyun Kim} 
\altaffiliation[Present address: ]{Quantum Technology Lab, SK Telecom, Seongnam-si, Gyeonggi-do, Korea 463-784}
\affiliation{Electrical and Computer Engineering Department, Duke University, Durham, N.C. 27708, USA}
\author{Peter Maunz}
\affiliation{Electrical and Computer Engineering Department, Duke University, Durham, N.C. 27708, USA}
\affiliation{Sandia National Laboratories, Albuquerque, N.M. 87123, USA}
\author{Jungsang Kim} \email{jungsang@duke.edu}
\affiliation{Electrical and Computer Engineering Department, Duke University, Durham, N.C. 27708, USA}

\begin{abstract}
Fast and efficient detection of the qubit state in trapped ion quantum information processing is critical for implementing quantum error correction and performing fundamental tests such as a loophole-free Bell test. In this work we present a simple qubit state detection protocol for a $^{171}$Yb$^+$ hyperfine atomic qubit trapped in a microfabricated surface trap, enabled by high collection efficiency of the scattered photons and low background photon count rate. We demonstrate average detection times of 10.5, 28.1 and 99.8\,$\upmu$s, corresponding to state detection fidelities of 99\%, 99.85(1)\% and 99.915(7)\%, respectively.
\end{abstract}

\maketitle

 Trapped ions provide a robust physical platform to realize quantum bits (qubits) for quantum information processing~\cite{WinelandJRNIST1998,BlattNature2008}. 
Effective qubit state detection can be achieved by driving a cycling transition between one of the qubit states and an auxiliary state with resonant light, causing state-dependent scattering~\cite{WinelandOL1980,NagourneyPRL1986,SauterPRL1986,BergquistPRL1986}. The state that scatters (does not scatter) light is commonly referred to as the {\em bright} ({\em dark}) state. The detection fidelity and speed are limited by undesired effects such as off-resonant optical pumping, background photon counts, and imperfect detection of the scattered photons. While the state detection fidelity has been improved using various strategies~\cite{SchaetzPRL2005,LangerThesis2006,HumePRL2007,BenhelmNatPhys2008,MyersonPRL2008},  long measurement times remain the dominant factor limiting the speed of trapped ion quantum information processing. By drastically improving the detection efficiency of the scattered photons, reducing the background photon count rates, and employing a time-tagging circuit, we demonstrate a substantial improvement in the qubit measurement time while maintaining a high fidelity. Fast qubit state detection is crucial for implementing quantum error correction~\cite{DiVincenzoPRL2007} and fundamental tests of quantum mechanics, such as a loophole-free Bell test~\cite{SimonPRL2003}.

The primary source of state detection error in direct hyperfine qubit measurements is off-resonant scattering which causes undesired conversion of the bright qubit state into the dark state, or vice versa. Other significant sources of error may include polarization impurity in the excitation beam~\cite{LangerThesis2006}, or overlap between the photon number distributions corresponding to dark and bright states. 
When the main contribution to the error is caused by one qubit state converting to the other, one can improve the fidelity by either shelving one of the qubit states to an auxiliary state that is highly unlikely to transition back to the qubit states, or by utilizing the arrival time of the scattered photons to identify the events where the qubit state went through a transition during the measurement~\cite{LangerThesis2006,MyersonPRL2008}. 
Another way to improve the fidelity is to extend the qubit measurement to multiple measurement attempts using a quantum logic gate~\cite{SchaetzPRL2005,HumePRL2007}. All demonstrated approaches suffer from low detection efficiency of photons scattered by the bright qubit state ($\approx 10^{-3}$), resulting in a slow measurement that limits efficient implementation of quantum error correction~\cite{DiVincenzoPRL2007}. Various strategies for implementing optical components integrated with the trapped ions to dramatically enhance the collection efficiency have been suggested~\cite{Noek2010,Shu2010,Streed2011,Merrill2011,MaiwaldPRA2012}, and can lead to improved detection speed and fidelity.

\begin{figure}[tb]
\begin{center}
\includegraphics[width=3.3in]{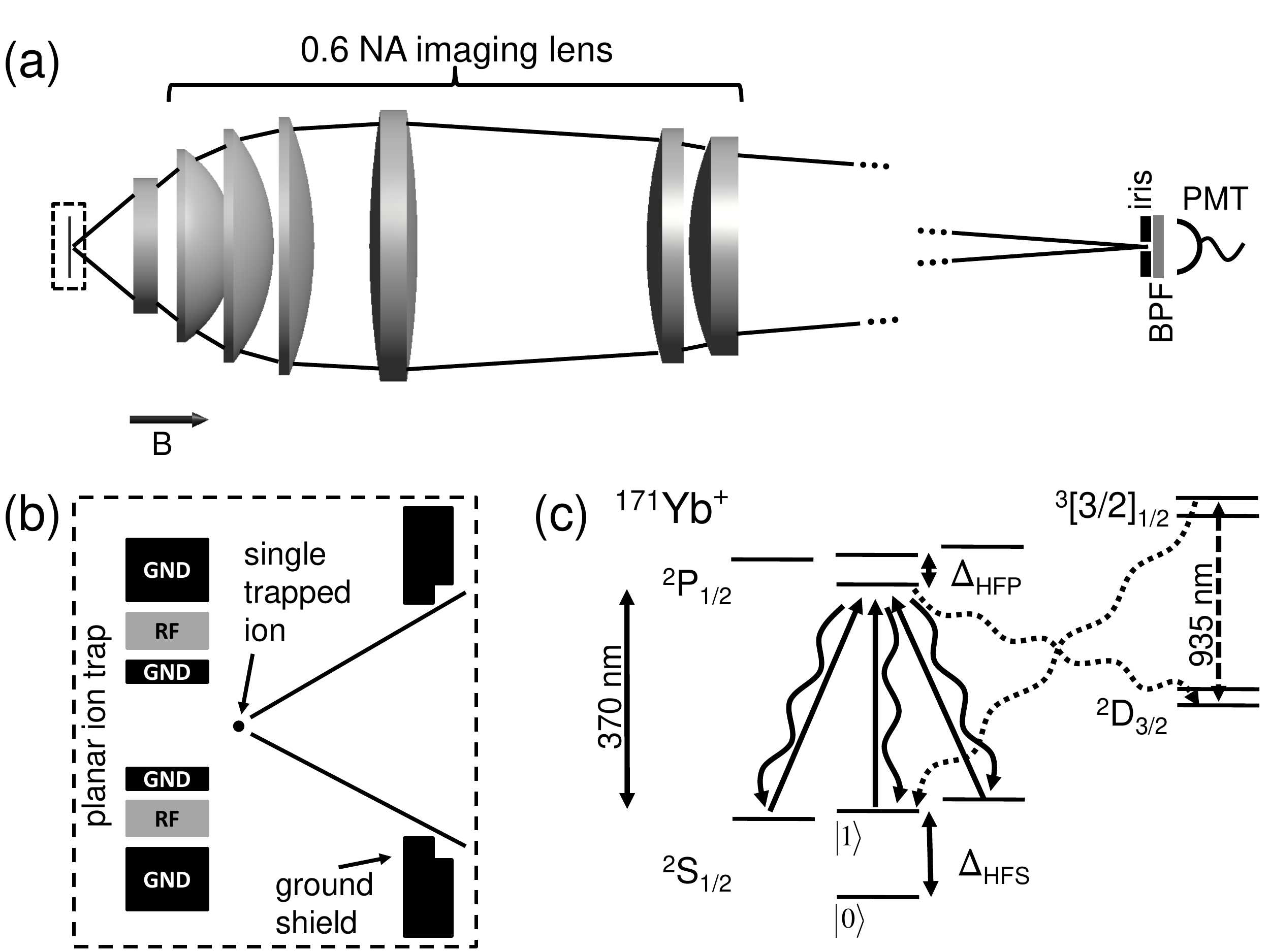}
\caption{(a) A single ion is trapped using a microfabricated surface RF trap and imaged through the vacuum window by a high numerical aperture (NA\,=\,0.6) lens. BPF: Band pass filter, PMT: Photomultiplier tube. The dashed-box region is magnified in (b) (not to scale). (c) The relevant energy levels in the $^{171}$Yb$^+$ ion used in the experiment. $\Updelta_\mathrm{HFS}$ and $\Updelta_\mathrm{HFP}$ indicate hyperfine splitting of the S and  P levels, respectively.}
\label{fig:system}
\end{center}
\end{figure}

Here, a single trapped $^{171}$Yb$^+$ ion is directly imaged using a custom objective lens (Photon Gear, Inc.) with a large object side numerical aperture (NA\,=\,0.6) capable of collecting 10\% of the total light scattered by the ion. The system is shown in Fig.~\ref{fig:system}(a). The ion trap and natural Yb oven are mounted in a standard six-inch ultra-high vacuum (UHV) octagonal chamber (Kimball Physics). 
A microfabricated radio frequency (RF) Paul trap (Sandia National Laboratories, Thunderbird) is used to trap a single ion 80\,$\upmu$m above the planar surface of the trap~\cite{StickarXiv2010,MonroeScience2013}. A ground plate is located 2\,mm above the trap surface, with a slot wide enough to accommodate NA\,$\approx$\,0.6 extending the length of the trap to allow imaging of the ion along the central linear trap axis (Fig.~\ref{fig:system}(b)). 
A custom re-entrant UHV window, anti-reflection coated for the ultraviolet light emitted by the ion, is used to position the inside surface of the window 10\,mm from the surface of the ion trap.  
Photons emitted by the ion are collected by the objective lens and focused on an iris used as a spatial filter. After the iris, the light passes through a 6\,nm  bandpass filter (90\% transmission, Semrock) and is detected with a photon counting photomultiplier tube (Ultra Bialkali PMT, Hamamatsu) with 32\% quantum efficiency at 370\,nm. The overall photon detection efficiency $\varepsilon$, was determined to be 2.2(1)\% by measuring the detected photon counts as a function of detection beam power for $^{174}$Yb$^+$, an isotope with no hyperfine structure that follows the simple scattering model of a two-level system. 

Figure \ref{fig:system}(c) shows the relevant energy levels of the hyperfine $^{171}$Yb$^+$ qubit. The qubit states $\ket{0}$ and $\ket{1}$ are defined as the two hyperfine states  $^2$S$_{1/2} \ket{F\!=\!0,m_f\!=\!0}$ and $^2$S$_{1/2} \ket{F\!=\!1,m_f\!=\!0}$, respectively. The ion is prepared in the $\ket{0}$ state by applying light resonant with $^2$S$_{1/2} \ket{F\!=\!1} \rightarrow \,^2$P$_{1/2} \ket{F\!=\!1}$ transition for 150\,$\upmu$s. 
An optional microwave $\uppi$-pulse resonant with the $^2$S$_{1/2}$ state hyperfine splitting (12.643 GHz) can rotate the ion to the $\ket{1}$ state with a $\uppi$-time of 240\,$\upmu$s. 

The state detection fidelity for the $^{171}$Yb$^+$ hyperfine qubit is measured by preparing the ion in either the $\ket{0}$ or $\ket{1}$ state, followed by a detection time during which the ion is exposed to the detection beam resonant with the $^2$S$_{1/2} \ket{F\!=\!1} \rightarrow \,^2$P$_{1/2} \ket{F\!=\!0}$ transition, focused to a 41\,$\upmu$m beam waist at the ion.  
In the absence of off-resonant scattering, an ion in the $\ket{0}$ (dark) state will scatter no photons, whereas an ion in the $\ket{1}$ (bright) state will experience a cycling transition (the $^2$S$_{1/2} \ket{F\!=\!0} \rightarrow \, ^2$P$_{1/2} \ket{F\!=\!0}$ transition is forbidden by dipole selection rules) and scatter photons at a rate of $R_\circ$. Throughout the initialization and detection process, a repump beam at 935 nm is used to prevent the ion from remaining in the dark $^2$D$_{3/2}$ state.

 In order to achieve the most efficient state detection, we require an optimal strategy to distinguish the bright state from the dark state by monitoring the scattered photons. 
The speed and fidelity of identifying the bright state increase with higher photon collection efficiency, while $R_\mathrm{dc}$, the sum of PMT dark counts (6.5\, Hz) and background photon counts (35\,Hz per 1\,$\upmu$W of detection beam power), degrades the fidelity of identifying the dark state. 
Detection error can also arise from the off-resonant scattering of the $\ket{1}$ ($\ket{0}$) state to the $^2$P$_{1/2} \ket{F\!=\!1}$ state, from which it can decay to the $\ket{0}$ ($\ket{1}$) state, at a rate $R_\mathrm{d}$ ($R_\mathrm{b}$). 

We experimentally determine all relevant scattering rates that impact the qubit state detection process. The scattering rate of the bright state $R_\circ$ takes into account optical pumping to coherent dark states~\cite{Berkeland2002}. Appropriate control of the detection beam polarization and Zeeman splitting $\delta$ destabilize these dark states, and the optimized scattering rate of the ion is given by:
\begin{equation}
R_{\circ,\mathrm{opt}} = \left(\frac{\Gamma}{6}\right) \frac{s_\circ}{1+\frac{2}{3} s_\circ + \left(\frac{2 \Delta}{\Gamma}\right)^2},
\end{equation}
where $\Gamma = 2 \pi \! \times \! 19.6$\,MHz is the linewidth of the $^2$P$_{1/2}$ state, $s_\circ = 2\Omega^2/\Gamma^2$ (with Rabi frequency $\Omega$) is the on-resonance saturation parameter, and $\Delta$ is the detuning of the detection beam from the cycling transition resonance. This result assumes an optimal Zeeman splitting of half the Rabi frequency ($\delta = \Omega/2$), which is a function of the optical power. In this work the Zeeman splitting is fixed (by fixing the magnetic field) at $\delta = 2 \pi \! \times \! 4.8$\,MHz. At higher powers the scattering rate decreases because the ion will pump into a coherent dark state on a timescale faster than the destabilization rate of the coherent dark states (given by $\delta$) and remains dark for a larger fraction of the time.

The rate at which an ion initially in the $\ket{1}$ state will pump to the $\ket{0}$ state is (for large detuning)
\begin{equation}
R_\mathrm{d} \approx \left( \frac{2}{3} \right) \left(\frac{1}{3}\right) \left( \frac{\Gamma}{2} \right) \left( \frac{2 \Omega^2}{\Gamma^2}\right) \left(\frac{ \Gamma}{2 \Delta_\mathrm{HFP}}\right)^2,
\end{equation}
where $\Delta_\mathrm{HFP}\!=\!2\pi\!\times\!2.1$\,GHz is the hyperfine splitting of the $^2$P$_{1/2}$ state. The factor of $2/3$ is due to the fact that one out of three states in the $^2$S$_{1/2} \ket{F\!=\!1}$ manifold is a coherent dark state at any given time, and so the expected reduction in the scattering rate for this transition is smaller than that for $R_\circ$. The factor of $1/3$ is the branching ratio of the $^2$P$_{1/2} \ket{F\!=\!1}$ states decaying into the dark $\ket{0}$ state. Similarly, the rate for $\ket{0}$ to pump into one of the bright $^2$S$_{1/2} \ket{F\!=\!1}$ states is
\begin{equation}
R_\mathrm{b} \approx \left( \frac{2}{3} \right) \left( \frac{\Gamma}{2} \right) \left( \frac{2 \Omega^2}{\Gamma^2}\right) \left(\frac{ \Gamma}{2 \left(\Delta_\mathrm{HFP}+\Delta_\mathrm{HFS}\right)}\right)^2,
\end{equation}
where $\Delta_\mathrm{HFS}\!=\!2\pi\times12.6$\,GHz is the hyperfine splitting of the  $^2$S$_{1/2}$ state. Here, the factor of $2/3$ is the branching ratio of the $^2$P$_{1/2} \ket{F\!=\!1}$ states decaying into the bright $^2$S$_{1/2} \ket{F\!=\!1}$ states. 

\begin{figure}[tb]
\begin{center}
\includegraphics[width=3.3in]{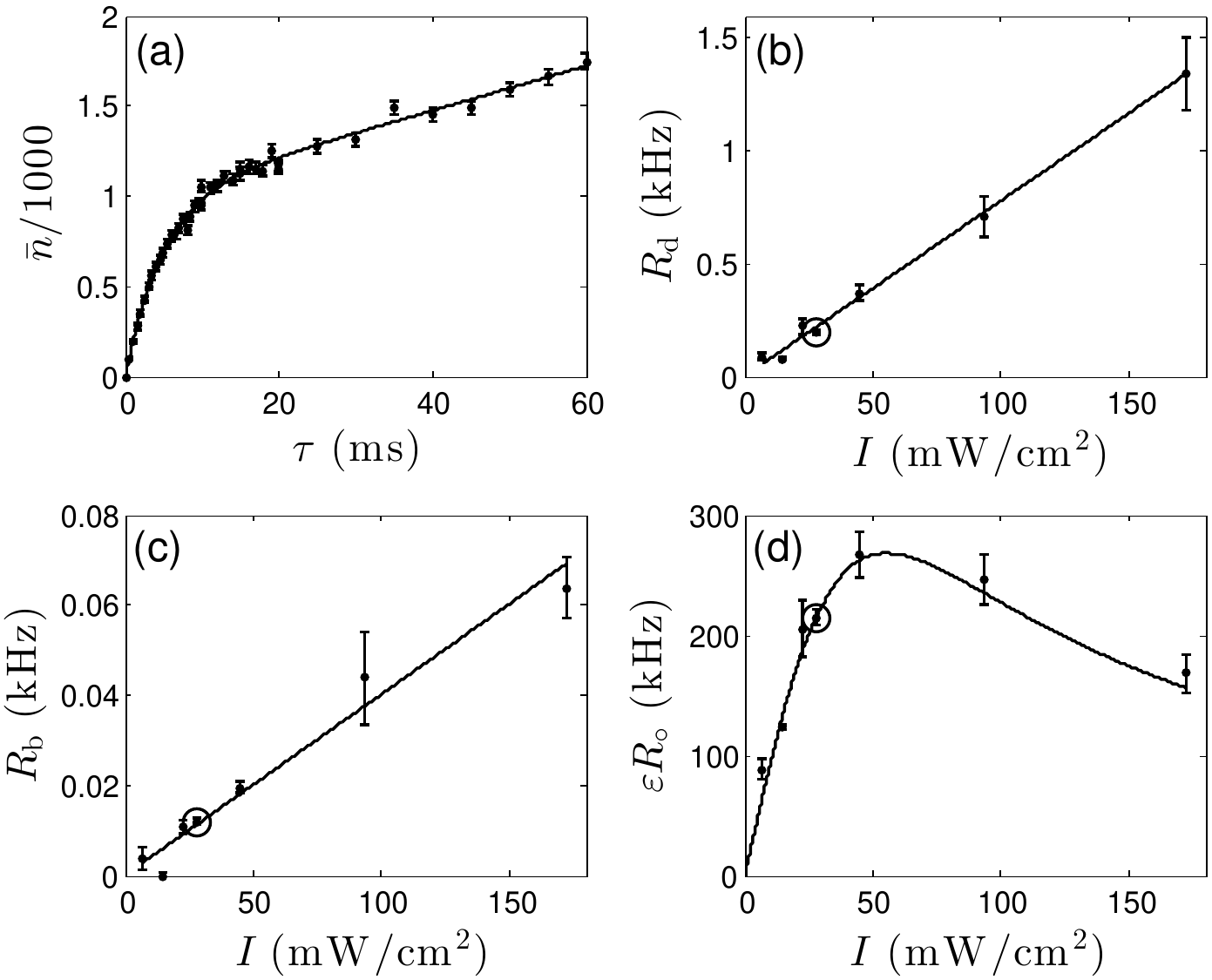}
\caption[Scattering, dark pumping, and bright pumping rate data vs. optical intensity]{(a) A sample plot of the average number of detected photons ($\bar{n}$) in a duration $\tau$ with a detection beam intensity of 29 mW/cm$^2$. (b) The dark pumping rate ($R_\mathrm{d}$), (c) the bright pumping rate ($R_\mathrm{b}$), and (d) the photon detection rate in the absence of dark pumping ($\varepsilon R_\circ$). $R_\mathrm{d}$ and $R_\mathrm{b}$ are fitted to a line in order to obtain the proportionality constant between the detection beam power and the square of the Rabi frequency, which is used to obtain the theoretical curve for $R_\circ$. Circled points in (b)-(d) correspond to the data shown in (a).}
\label{fig:rates}
\end{center}
\end{figure}

These scattering rates are measured by preparing the ion in the $\ket{1}$ state and detecting the scattered photons over a duration $\tau$. We fit this data to the function 
\begin{equation}
\bar{n}(\tau) = \int_0^\tau \! \varepsilon R_\circ p_1(t) \, \mathrm{d}t,
\end{equation}
which is obtained by simultaneously solving the equations $\dot{p}_1 = R_\mathrm{b} p_0 - R_\mathrm{d} p_1$ and $p_0 + p_1 = 1$ for $p_1(t)$, where $p_1(t)$ ($p_0(t)$) is the probability of the ion to be in the bright states (dark state) as a function of time (an example is shown in Fig.~\ref{fig:rates}(a)). From these measurements, we can experimentally determine the dark pumping rate $R_\mathrm{d}$, bright pumping rate $R_\mathrm{b}$ and the photon detection rate $\varepsilon R_\circ$ as a function of optical detection beam power (Fig.~\ref{fig:rates}(b), ~\ref{fig:rates}(c), and ~\ref{fig:rates}(d), respectively).

One strategy to discriminate the bright and dark states is to count the number of detected photons over a given detection period $\uptau_\mathrm{max}$ and compare it to a fixed threshold. 
Alternatively, if the photon time of arrival information is available, one can utilize a more effective decision procedure to reduce the average detection time without compromising the detection fidelity~\cite{Myerson2008}. For the conditions in our experiments with high photon detection efficiency and low background counts, we attribute any detection events with zero photons as dark states and those with two or more photons as bright states. The ambiguity arises for those events where the PMT registers one photon during the detection period: these counts can arise from either a background count for a dark qubit state, or from a bright qubit state that pumps dark after one photon is detected.

We employ a fast detection scheme where we monitor the arrival time of at most the first two photons. If the first photon arrives before a cut-off time $\uptau_\mathrm{c}$, the photon more likely originated from a bright state, and the state is determined to be bright.  
However, if the first photon arrives after $\uptau_\mathrm{c}$, we wait to see if a second photon arrives before a maximum wait time $\uptau_\mathrm{max}$. The state is declared bright at the arrival time of a second photon, whereas if no more photons are detected, the first photon is more likely from a background count and the state is determined to be dark.
  
For an ion in the $\ket{0}$ state, the probability of obtaining the first PMT detection event on or before a time $t$ grows linearly and depends only on $R_\mathrm{dc}$ ($P_0(t) \approx R_\mathrm{dc} t$). The probability of detecting the first photon on or before $t$ for an ion in the $\ket{1}$ state is approximately given by: 
\begin{equation}
P_1(t) \approx R_\mathrm{d}/(\varepsilon R_\circ) \left[1-\exp(-\varepsilon R_\circ t) \right],
\end{equation}  
where the second factor is the probability of obtaining a non-zero number of PMT events before $t$ is reached and the first factor is the probability of off-resonant scattering to $\ket{0}$ after obtaining a single photon (in the limit where $\varepsilon R_\circ t \! \gg \!1$ and therefore the Poissonian probability of obtaining only a single photon is negligible).  
By solving for the time where $P_1(t) - P_0(t)$ is maximized, we obtain 
\begin{equation}
\uptau_\mathrm{c} = \ln\left(R_\mathrm{d}/R_\mathrm{dc}\right)/\varepsilon R_\circ,
\end{equation}
before which the first photon detection event is more likely to have originated from a bright state. The maximum time $\uptau_\mathrm{max}$ we are willing to wait is determined experimentally. The overall error probability starts to increase after $\uptau_\mathrm{max}$ due to the small chance of off-resonant pumping of the bright states to $\ket{0}$ before a photon is collected, and also from the increasing error probability for ions in $\ket{0}$ caused by non-zero dark count and background scattering rates.  The overall error probability reaches a minimum at $t = \uptau_\mathrm{max}$, and photon detection events after this time are ignored.  The dark state decision will always take the maximum time $\uptau_\mathrm{max}$, but the bright state decision typically happens much more quickly as the first or second photon is detected, reducing the average state detection time to well below $\uptau_\mathrm{max}$. 

\begin{figure}[tb]
\begin{center}
\includegraphics[width=3.3in]{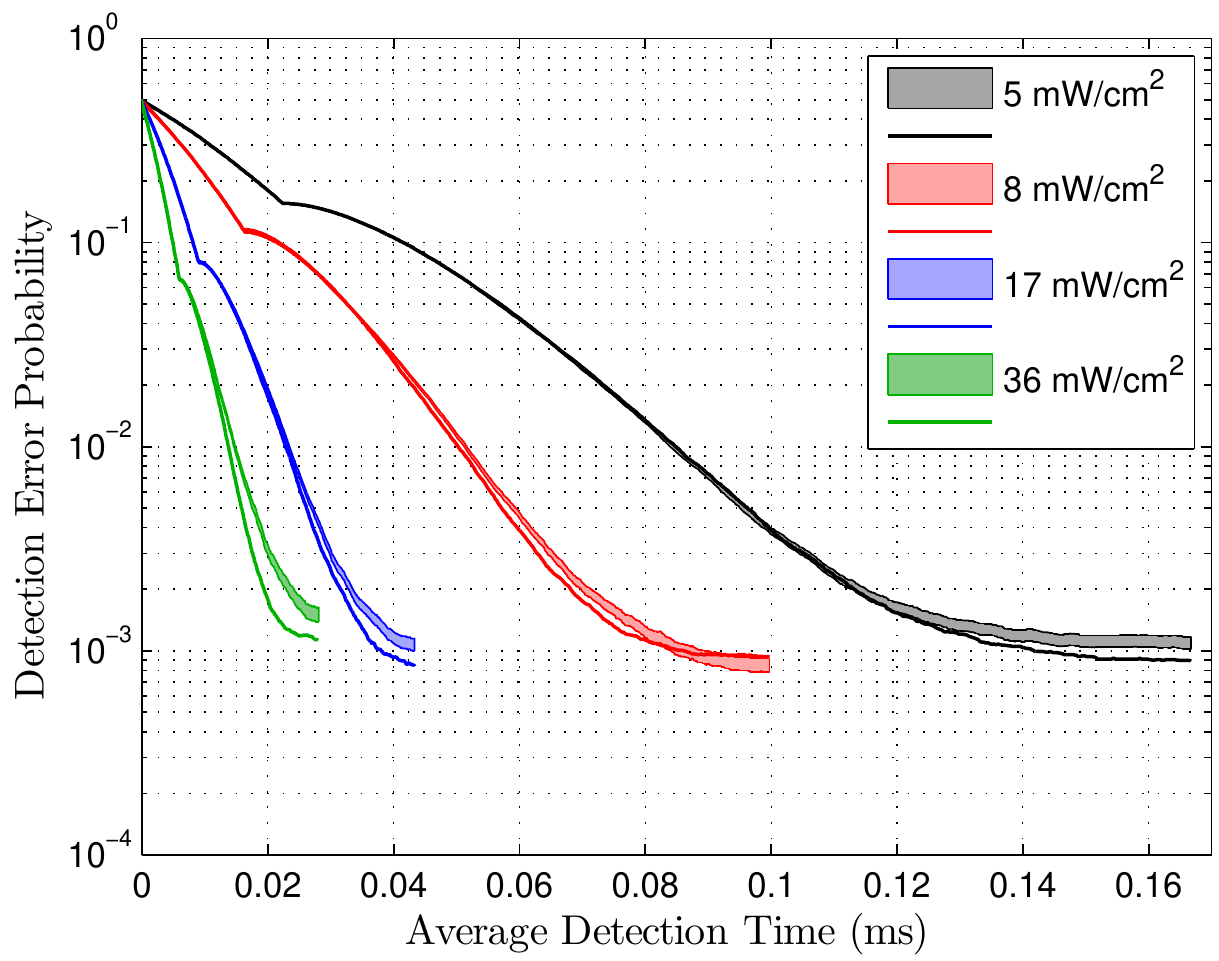}
\caption{(color online) State detection experimental results and simulations for different optical intensities of the detection beam. The wide translucent trends are the results from 100,000 experiments where the width indicates the $1/\mathrm{e}$ confidence interval, and the solid lines are simulation results. 
}
\label{fig:statedetectiondata}
\end{center}
\end{figure}

The average state detection error determined from 100,000 experiments (50,000 each for $\ket{0}$ and $\ket{1}$ prepared states) is shown in Fig.~\ref{fig:statedetectiondata}, as a function of the average detection time for various levels of detection beam intensities. 
The curves are generated by plotting the average result from all samples for a maximum detection time ranging from zero to $\uptau_\mathrm{max}$, and the sharp bends correspond to the detection time reaching $\uptau_\mathrm{c}$. 
The solid lines are generated from the simulation using the relevant rates determined experimentally (procedures shown in Fig.~\ref{fig:rates}), and match very well with the experimental results. 
The  results indicate experimental qubit state detection fidelities of 99.85(1)\% after an average detection time of 28.1\,$\upmu$s (worst case 51.4\,$\upmu$s, intensity 36\,mW/cm$^2$), and 99.915(7)\% after an average detection time of 99.8\,$\upmu$s (worst case 181.6\,$\upmu$s, intensity 8\,mW/cm$^2$). 
For applications where higher qubit detection speeds are desired at the expense of reduced fidelity, we can use the arrival of the first photon to classify the state to be $\ket{1}$, and dramatically reduce $\uptau_\mathrm{max}$. Using this approach, our system is capable of achieving a 99\% detection fidelity for $\uptau_\mathrm{max} = 17.0$\,$\upmu$s, with an average detection time of 10.5\,$\upmu$s.

We have demonstrated a substantial reduction in the direct qubit state detection time for a $^{171}$Yb$^+$ hyperfine qubit using high NA optics while maintaining a low error rate. Our high photon collection efficiency and low background count rate allow us to realize a simple state detection protocol using time-of-arrival information for the first two detected photons. The fidelity of the direct state detection for our hyperfine qubit is ultimately limited by off-resonant scattering of the qubit states. Further fidelity improvement can be achieved using a shelving technique to reduce error originating from the $\ket{1}$ state pumping to the dark $\ket{0}$ state~\cite{LangerThesis2006,BenhelmNatPhys2008,MyersonPRL2008}.  Other strategies that increase the photon collection efficiency can be used in tandem with our approach to increase both fidelity and detection speed~\cite{Noek2010,Shu2010,Merrill2011,MaiwaldPRA2012}.

 The average detection time demonstrated in this work is comparable to the timescale over which a two-qubit gate is typically performed in trapped ion systems. This shows that the state detection might not be the rate-limiting procedure in quantum computations which employ error correction~\cite{DiVincenzoPRL2007}, and also reduces the distance required for realizing space-like separation in a loophole-free Bell test to well below 10\,km, making an ion trap system a feasible candidate for this task~\cite{SimonPRL2003}.

This work was supported by the Office of the Director of National Intelligence and Intelligence Advanced Research Projects Activity through the Army Research Office.


\begin{thebibliography}{26}

\bibitem{WinelandJRNIST1998}
   D. J. Wineland, C. Monroe, W. M. Itano, {\it et al.},
   {J. Res. Natl. Inst. Stand. Technol.} {\bf 103}, 259 (1998).

\bibitem{BlattNature2008}
   R. Blatt and D. J. Wineland,
   {Nature} {\bf 453}, 1008 (2008).

\bibitem{WinelandOL1980}
   D. J. Wineland, J. C. Bergquist, W. M. Itano and R. E. Drullinger, 
   {Opt. Lett.} {\bf 5}, 245 (1980).

\bibitem{NagourneyPRL1986}
   W. Nagourney, J. Sandberg and H. Dehmelt, 
   {Phys. Rev. Lett.} {\bf 56}, 2797 (1986).

\bibitem{SauterPRL1986}
   Th. Sauter, W. Neuhauser, R. Blatt and P. E. Toschek,
   {Phys. Rev. Lett.} {\bf 57}, 1696 (1986).

\bibitem{BergquistPRL1986}
   J. C. Bergquist, R. G. Hulet, W. M. Itano and D. J. Wineland,
   {Phys. Rev. Lett.} {\bf 57}, 1699 (1986).

\bibitem{SchaetzPRL2005}
   T. Schaetz, M. D. Barrett, D. Leibfried {\it et al.},
   {Phys. Rev. Lett.} {\bf 94}, 010501 (2005).

\bibitem{LangerThesis2006}
   C. E. Langer,
   {Ph.D. Thesis, University of Colorado at Boulder} (2006).

\bibitem{HumePRL2007}
   D. B. Hume, T. Rosenband and D. J. Wineland,
   {Phys. Rev. Lett.} {\bf 99}, 120502 (2007).

\bibitem{BenhelmNatPhys2008}
   J. Benhelm, G. Kirchmair, C. F. Roos and R. Blatt,
   {Nature Phys.} {\bf 4}, 463 (2008).

\bibitem{MyersonPRL2008}
   A. H. Myerson, D. J. Szwer, S. C. Webster {\it et al.},
   {Phys. Rev. Lett.} {\bf 100}, 200502 (2008).

\bibitem{Olms2007}
    S. Olmschenk, K. C. Younge, D. L. Moehring {\it et al.},
    {Phys. Rev.} {\bf A 76},  052314 (2007).

\bibitem{DiVincenzoPRL2007}
   D. P. DiVincenzo and P. Aliferis,
   {Phys. Rev. Lett.} {\bf 98}, 020501 (2007).

\bibitem{SimonPRL2003}
   C. Simon and W. T. M. Irvine,
   {Phys. Rev. Lett.} {\bf 91}, 110405 (2003).

\bibitem{Noek2010}
    R. Noek, C. Knoernschild, J. Migacz {\it et al.},
    {Opt. Lett.} {\bf 35}, 2460 (2010).

\bibitem{Shu2010}
    G. Shu, N. Kurz, M. R. Dietrich and B. B. Blinov,
    {Phys. Rev. A} {\bf  81}, 042321 (2010). 
    
\bibitem{Streed2011}
    E. W. Streed, B. G. Norton, A. Jechow {\it et al.},
    {Phys. Rev. Lett.} {\bf 106}, 010502 (2011).

\bibitem{Merrill2011}
     J. T. Merrill, C.  Volin, D. Landgren {\it et al.},
    {New J.  Phys.} {\bf 13} 103005, (2011).

\bibitem{MaiwaldPRA2012}
   R. Maiwald, A. Golla, M.  Fischer {\it et al.},
   {Phys. Rev. A} {\bf 86}, 043431, (2012).

\bibitem{StickarXiv2010}
   D. Stick, K. M. Fortier, R. Haltli {\it et al.},
   arXiv:1008.0990 (2010).
   
\bibitem{MonroeScience2013}
   C. Monroe and J. Kim,
   {Science} {\bf 339}, 1164 (2013).

\bibitem{DiVincenzoCriteria}
    D. DiVincenzo and D. Loss,
    {Superlattices and Microstructures} {\bf 23}, 419 (1998).

\bibitem{Moehring2007}
    D. L. Moehring, P. Maunz, S. Olmschenk {\it et al.},
    {Nature} {\bf 449}, 68 (2007).

\bibitem{Trupke2005}
    M. Trupke, E. A. Hinds, S. Eriksson {\it et al.},
    {Appl. Phys. Lett.} {\bf 87} 211106 (2005).

\bibitem{Herskind2011}
    P. F. Herskind, S. X. Wang, M. Shi {\it et al.},
    {Opt. Lett.} {\bf 36} 3045 (2011).

\bibitem{Myerson2008}
    A. H. Myerson, D. J. Szwer, S. C. Webster {\it et al.},
    {Phys. Rev. Lett.} {\bf 100}, 200502 (2008).

\bibitem{Berkeland2002}
    D. J. Berkeland and M. G. Boshier,
    {Phys. Rev. A} {\bf 65}, 033413 (2002).

\bibitem{Kim2011}
    T. Kim, P. Maunz and J. Kim,
    {Phys. Rev. A} {\bf 84}, 063423 (2011).

%
%
%
%
%
%
%
%
%
%
%
%
%
%
%
%
%
%
%
%
%
%
%
%
%
%

\end{thebibliography}
\end{document}